# Optimality of mutation and selection in germinal centers

Jingshan Zhang and Eugene I. Shakhnovich

*Department of Chemistry and Chemical Biology, Harvard University, 12 Oxford Street, Cambridge, MA 02138*

**February 7, 2010**

**Abstract:** The population dynamics theory of B cells in a typical germinal center could play an important role in revealing how affinity maturation is achieved. However, the existing models encountered some conflicts with experiments. To resolve these conflicts, we present a coarse-grained model to calculate the B cell population development in affinity maturation, which allows a comprehensive analysis of its parameter space to look for optimal values of mutation rate, selection strength, and initial antibody-antigen binding level that maximize the affinity improvement. With these optimized parameters, the model is compatible with the experimental observations such as the ~100-fold affinity improvements, the number of mutations, the hypermutation rate, and the "all or none" phenomenon. Moreover, we study the reasons behind the optimal parameters. The optimal mutation rate, in agreement with the hypermutation rate *in vivo*, results from a tradeoff between accumulating enough beneficial mutations and avoiding too many deleterious or lethal mutations. The optimal selection strength evolves as a balance between the need for affinity improvement and the requirement to pass the population bottleneck. These findings point to the conclusion that germinal centers have been optimized by evolution to generate strong affinity antibodies effectively and rapidly. In addition, we study the enhancement of affinity improvement due to B cell migration between germinal centers. These results could enhance our understandings to the functions of germinal centers.

**Author Summary:** The antibodies in our immune system could efficiently improve their abilities in recognizing new antigens. This is done with the help of proliferation, mutation and selection of B cells which carry antibodies; but we have difficulties in developing a quantitative description of this adaptation process which is consistent with the various aspects of experimental observations. Based on the knowledge from experiments, here we present a theoretical model to calculate the numbers of B cells with different antigen recognizing abilities all the time, and look for the best possible design that improves the antigen recognizing ability most efficiently. We find that the best possible design is consistent with the experimental observations, pointing to the conclusion that the immune system has been optimized in evolution. We then study the trade-offs leading to the optimization of the design. The results will not only improve our understanding of the functions in immune system, but also reveal the design principles behind the details. In addition, the study enhances our understanding to the population dynamics in evolution.

## Introduction

As one of the adaptive immune responses[1,2,3,4], Affinity Maturation (AM) is the procedure in germinal centers (GC) to develop Immunoglobulins (Ig), i.e., antibodies, with increased affinities to a new antigen. Understanding the basic functional and physical principles of GC kinetics is not only important in medical science, but also contributes to the fundamental understanding of molecular evolution[5]. Some mathematical models[6,7] made an important effort to describe the B cell population in a typical GC as a result of dynamic interactions between mutation and selection. This effort, complementary to the studies using Ig sequence data (e.g. [8]), is very important in revealing the functions of germinal centers. However, there are still two puzzles in the models. First, the experimentally observed somatic hypermutation[9] rate seems so high that even B cells expressing antibodies with improved affinities are easily spoiled by the majority of deleterious mutations. To resolve the conflict, it is proposed[6,10] that the mutation could be switched off periodically. Second, even if the mutation rate is tuned periodically, the calculated affinity improvement (up to 15-fold) is still not comparable with the observed improvements (~100-fold[11,12,13,14]). Third, a further analysis indicated that the models did not reproduce the ''all or none'' phenomenon[8].

Since most of the parameters in the models are estimated from experiments with considerable uncertainties, it is possible to reconcile the discrepancies by revising the parameter values. However, the number of parameters is not small, and it is unclear how to find the best parameter values. We notice that the affinity dependent selection results from antigen binding kinetics, salvation, and recirculation of B cells. If we replace these steps by a phenomenological linear function of affinity to represent the selection, the calculations will be significantly simplified, and it will be possible to explore the parameter space and look for the optimal design. We hypothesize that AM has been optimized in evolution, and expect the B cell population dynamics with these optimized parameters to reproduce realistic AM.

## Methods

**The model.** Each GC is believed to start from a few precursor B cells [15]. These precursor B cells first replicate at the perimeter of follicles[16], and the number of B cells in a GC reaches thousands when mutations and affinity dependent selections are turned on [7,17,18,19,20]. Our model is constructed as follows:

First, the model describes the stage with mutation and selection, where the total number of germline (initial) B cells in the hundreds of GC in a spleen is $\sim 10^6$. B cells replicate at 3~4divisions/day[21,22,23]. To be specific we use 4divisions/day in our calculations, corresponding to exponential birth rate $r=4\ln 2=2.8$/day. Upon every division about one of the two daughter cells are mutated[9], so we find the total mutation rate $m_{total} \approx 2.8 / day$ from $\exp(-m_{total} 0.25) \approx 1 - 1/2$, although in the calculation below we will explore different values of $m_{total}$ to search for the optimal design. It is estimated[24,25] that about 50% mutations are silent, 30% are lethal, and the rest 20% are the affinity-affecting mutations. The lethal mutation rate $0.3 m_{total}$ is an effective death rate. After taking care of the lethal mutations and neglecting silent mutations, we concentrate on the affinity-affecting mutations. The distribution $W(\Delta X)$ of affinity change $\Delta X$ upon such mutations (Fig. 1) is estimated from the protein interaction (PINT) database[26] (see Supporting Information for details). Note that only 4.9% of affinity-affecting mutations (equivalent to 1% of all mutations) improve the affinity.

Second, selection is considered on the basis of the recent two-photon spectroscopic studies which indicate B cells undergo multiple rounds of mutation and selection[27] through migration within GC[3,4], consistent with the recycling hypothesis[6,7]. B cells compete for antigens and salvation from apoptosis, therefore the probability for a B cell to survive in each round depends on the probability to bind an antigen in the round, which in turn depends on the affinity of the B cell's Igs to the antigens. According to a standard Langmuir adsorption isotherm the probability to bind an antigen in a round of selection is $K_a \cdot C_A / (K_a \cdot C_A + 1)$ where the association constant $K_a$ (in units of 1/M) describes the affinity and $C_A$ is the antigen concentration on the follicular dendritic cells of GCs. If we describe the affinity by the binding free energy $X = kT \ln(1/K_a)$, the probability for B cells to survive at each round is $\exp[-(X - X_0)/kT]$ for weak affinity, $X > X_0$, where $X_0 = kT \ln(C_A)$; and saturates at 1 for strong affinity, $X < X_0$, reflecting the observation of an affinity threshold[28]. Define $\tau$ as the typical time scale for each recycling round. Selection scales the population size as $\exp[-(t/\tau)(X - X_0)/kT]$ after time t for weak affinity B cells, so death/apoptosis rate is $(X - X_0)/(kT\tau)$.

Finally, including the replications, lethal mutations, and selections discussed above, the effective growth rate of the B cell population B(X) is

$$B(X) = \begin{cases} r - 0.3 m_{total} - b(X - X_0) & X > X_0 \quad (1a) \\ r - 0.3 m_{total} & X < X_0 \quad (1b) \end{cases} \quad (1)$$

where the selection strength $b = 1/(kT\tau)$ comes from the above discussion. Defining $X^*$ and $K_a^*$ as the "neutral" affinity, $B(X^*) = 0$, we find $X^* = X_0 + (r - 0.3 m_{total})/b$ from Eq. (1a). Population decrease is expected for weak affinity $X > X^*$, and $X^*$ is controlled by $X_0$ or $C_A$, which is in turn determined by antigen density, reflecting an antigen dosage effect on AM.

**Differential equation and analytic solution.** Now we are ready to write the mean-field differential equation for the population of B cells presenting Ig with affinity $X$ at time $t$:

$$\frac{dN(X,t)}{dt} = B(X)N(X,t) + m\int W(X - X')N(X',t)dX' - mN(X,t) \quad (2)$$

where $m \equiv 0.2 m_{total}$ is rate of the affinity-affecting mutations. (A similar equation was considered, in a different context and for fixed population size by Tsimring et. al. [29]) As mentioned above, replications, lethal mutations and selections are included in *B(X)*, and the affinity-affecting mutations are characterized by $W(\Delta X)$ in Fig. 1, where $\Delta X = X - X'$ is the affinity change.

We can solve the mean field equation (2) when Eq. (1) can be simplified as Eq. (1a), i.e. the effective growth rate (i.e. replication rate minus death rate) depends linearly on the binding free energy in the whole range. Then we rewrite growth rate $B(X) = b_0 - bX$ where $b_0 \equiv r - 0.3m_{total} + bX_0$. We introduce a Fourier transform of the population:

$$N(t,\omega) = \frac{1}{2\pi} \int N(X,t) e^{-i\omega X} dX \qquad (3)$$

Then in terms of the Fourier transform population, Eq. (2) looks simple:

$$\frac{\partial N(\omega,t)}{\partial t} = b_0 N(\omega,t) - ib \frac{\partial N(\omega,t)}{\partial \omega} + m[2\pi W(\omega) - 1]N(\omega,t) \qquad (4)$$

where

$$W(\omega) = \frac{1}{2\pi} \int e^{-i\omega \cdot \Delta X} W(\Delta X) d\Delta X$$

Now denoting

$$Q = \ln N(\omega,t) \qquad (5)$$

we get

$$\frac{\partial Q}{\partial t} + ib \frac{\partial Q}{\partial \omega} = b_0 + m[2\pi W(\omega) - 1] \qquad (6)$$

We seek solution of this equation in the form:

$$Q = R(\omega - ibt) + S(\omega) \qquad (7)$$

and get for S:

$$ib \frac{\partial S}{\partial \omega} = b_0 + m(W(\omega) - 1) \qquad (8)$$

or

$$S = i\frac{b_0 - m}{-b}\omega - i\frac{2\pi m}{b} \int_{-\infty}^{\omega} W(\omega')d\omega' + C \qquad (9)$$

where C is constant to be determined.

The function R and the constant C are determined from the initial condition at t=0, i.e., the germline distribution of affinities. The general results applicable for different initial conditions will be addressed shortly after; for the moment we adopt a most common initial condition, Gaussian distribution

$$N(X,t=0) = \frac{N_0}{\sqrt{2\pi\Sigma^2}} \exp\left[-\frac{(X-X_{av})^2}{2\Sigma^2}\right] \qquad (10)$$

of affinity, or in Fourier space:

$$N(\omega,0) = \frac{N_0}{2\pi} \exp\left[-\frac{1}{2}\Sigma^2 \omega^2 - i\omega X_{av}\right]$$

Now the function R and constant C can be easily determined from the relation:

$$N(\omega,0) = e^{Q(\omega,0)}$$

which immediately gives us $C = \ln(N_0/2\pi)$ and final result for R (and Q):

$$R(\omega) = -\frac{1}{2}\Sigma^2 \omega^2 - i\omega X_{av} + i\frac{b_0 - m}{b}\omega + i\frac{2\pi m}{b} \int_{-\infty}^{\omega} W(\omega')d\omega'$$

Plugging these results to Eq. (7), we obtain the solution

$$Q(\omega,t) = -\frac{1}{2}\Sigma^2\omega^2 - i\omega[X_{av} - \Sigma^2 bt] + t(b_0 - m - bX_{av}) + \frac{1}{2}b^2\Sigma^2 t^2 + i\frac{2\pi m}{b}\int_\omega^{\omega+ibt}W(\omega')d\omega' + \ln\frac{N_0}{2\pi}$$

and accordingly

$$N(\omega,t) = \frac{N_0}{2\pi}\exp\left[-\frac{1}{2}\Sigma^2\omega^2 - i\omega(X_{av} - \Sigma^2 bt) + t(b_0 - m - bX_{av}) + \frac{1}{2}b^2\Sigma^2 t^2 + i\frac{2\pi m}{b}\int_\omega^{\omega+ibt}W(\omega')d\omega'\right]$$

(11)

Noticing

$$2\pi\int_\omega^{\omega+ibt}W(\omega')d\omega' = \int_\omega^{\omega+ibt}d\omega'\int d\Delta X W(\Delta X)e^{-i\omega'\Delta X} = \int d\Delta X W(\Delta X)\int_\omega^{\omega+ibt}d\omega' e^{-i\omega'\Delta X},$$

$$= \frac{1}{-i}\int d\Delta X \frac{W(\Delta X)}{\Delta X}e^{-i\omega\Delta X}(e^{-bt\Delta X} - 1)$$

we can define a Fourier transform pair $f(\Delta X, t) = \frac{2\pi m}{b}\frac{W(\Delta X)}{\Delta X}(e^{-bt\Delta X} - 1)$ and $F(\omega, t) = \exp\left[\frac{m}{-b}\int d\Delta X \frac{W(\Delta X)}{\Delta X}e^{-i\omega\Delta X}(e^{-bt\Delta X} - 1)\right]$, and another Fourier transform pair

$$g(X,t) = \frac{N_0}{\sqrt{2\pi\Sigma^2}}\exp\left[(b_0 - m - bX_{av})t + \frac{1}{2}\Sigma^2 b^2 t^2 - \frac{(X - X_{av} + \Sigma^2 bt)^2}{2\Sigma^2}\right]$$

and $G(\omega,t) = \frac{N_0}{2\pi}\exp\left[-\frac{1}{2}\Sigma^2\omega^2 - i\omega(X_{av} - \Sigma^2 bt) + t(b_0 - m - bX_{av}) + \frac{1}{2}b^2\Sigma^2 t^2\right]$, then Eq. (11) becomes

$$N(\omega,t) = G(\omega,t)\exp[F(\omega,t)] = G(\omega,t)\left[1 + F(\omega,t) + \tfrac{1}{2!}F^2(\omega,t) + \tfrac{1}{3!}F^3(\omega,t) + \cdots\right] \quad (12)$$

With help of the Fourier convolution theorem

$$\frac{1}{2\pi}\int dX e^{-i\omega\cdot\Delta X}\left[\int f(X - X')g(X')dX'\right] = 2\pi\cdot F(\omega)G(\omega),$$

we obtain the final EXACT solution for the selection-diffusion equation (2)

$$N(X,t) = g(X,t) + \frac{1}{2\pi}\int g(X')f(X - X')dX' + \frac{1}{2!}\frac{1}{(2\pi)^2}\iint g(X')f(X''-X')f(X - X'')dX'dX'' + \ldots$$

$$= g(X,t) + \frac{1}{2\pi}\int g(X - \Delta X)f(\Delta X)d\Delta X + \frac{1}{2!}\frac{1}{(2\pi)^2}\iint g(X - \Delta X_1 - \Delta X_2)f(\Delta X_1)f(\Delta X_2)d\Delta X_1 d\Delta X_2 + \cdots$$

$$= \tilde{N}(X,t)e^{-mt} + \sum_{j=1}^\infty \frac{1}{j!}\left(\frac{m}{-b}\right)^j\int d\Delta X_1 \frac{W(\Delta X_1)}{\Delta X_1}(e^{-bt\Delta X_1} - 1)\cdots\int d\Delta X_j \frac{W(\Delta X_j)}{\Delta X_j}(e^{-bt\Delta X_j} - 1)\tilde{N}(X - \Delta X_1 - \cdots - \Delta X_j, t)e^{-mt}$$

(13)

where $\tilde{N}(X,t) = g(X,t)e^{mt} = N(X, t=0)\exp[(b_0 - bX)t]$ is the population profile in the limit of *m=0*. Due to the linearity of Eq. (2), Eq. (13) applies generally to any initial condition. The total population is

$$N(t) = \int N(X,t)dX = \int N(X,t)e^{-i\cdot 0\cdot X}dX = 2\pi \cdot N(\omega=0,t)$$

$$= N_0 \exp\left[(b_0 - bX_{av})t + \frac{1}{2}\Sigma^2 b^2 t^2\right] \cdot \exp\left[-mt - \frac{m}{b}\int \frac{(e^{bt\cdot \Delta X}-1)}{\Delta X}W(\Delta X)d\Delta X\right] \quad (14)$$

$$= \tilde{N}(t)\exp\left[-mt - \frac{m}{b}\int \frac{(e^{bt\cdot \Delta X}-1)}{\Delta X}W(\Delta X)d\Delta X\right]$$

The effects of mutations are in the exponent containing $m$, and this exponent applies universally to initial conditions other than Eq. (10). At long enough time beneficial mutations $\Delta X < 0$ contribute a growing term $\frac{m}{-b}\int_0^\infty \frac{e^{bt\Delta X}}{\Delta X}W(\Delta X)d\Delta X$ because $e^{-bt\Delta X} \gg 1$; while the deleterious mutations $\Delta X > 0$ mainly contribute to the population reduction term $-mt$ because $e^{-bt\Delta X} \ll 1$.

**Intuitive derivation.** An intuitive derivation of (14) helps us understand their physical meaning. For convenience we use discrete values of $\Delta X_i = -ih$ in Fig. 1, and write the total mutation rate as a sum of individual mutation rates, $m = \sum_i m_i$ where $m_i$ is the rate of mutation that change affinity by $\Delta X_i = -ih$. Then Eq. (14) becomes:

$$N(t) = \tilde{N}(t)\prod_i \exp[-m_i t + \frac{m_i}{bih}(e^{biht}-1)]. \quad (15)$$

Note that beneficial mutations ($i>0$) contribute $-m_i t + \frac{m_i}{bih}(e^{biht}-1) \xrightarrow{t\to\infty} \frac{m_i}{bih}e^{biht}$ to $\ln N(t)$. Therefore in the long time limit the beneficial mutations with leads to a super-exponential population growth $N(t) \sim \exp[\exp(t)]$, in comparison to the exponential population growth without mutation. If we call the strongest affinity in the population as the "nose" affinity, as the "nose" moves toward stronger affinity, the subpopulation at the "nose" grows faster, and the speed of "nose" movement becomes faster.

The effects of different mutations in Eq. (15) can be factored out, because a) the rate $m_i$ for a mutation of $\Delta X_i = -ih$ to emerge is independent of the affinity distribution $N(X,t)$, and b) the contribution of a series of mutations $\{\Delta X_i\}$ occurring at $\{t_i\}$ to the subpopulation size at a later time $t$ is $\prod_i \exp[-b\Delta X_i(t-t_i)]$, a product of individual mutation factors. Therefore, to understand Eq. (14), we can focus on effect of only one mutation type with affinity change $\Delta X_i = -ih$ and individual mutation rate $m_i$.

First, in the limit of $b \to 0$ in Eq. (1), the benefit of mutations is turned off, and the average number of mutations B cells experience in the interval $t$ is $m_i t$, and the probability to experience $j$ mutations is a standard Poisson distribution $P_j(t) = \exp(-m_i t)\frac{(m_i t)^j}{j!}$. It is straight forward to verify

$$N(t) = \tilde{N}(t)\sum_j P_j(t) = \tilde{N}(t)\exp(-m_i t)\sum_{j=0}^\infty \frac{1}{j!}(m_i t)^j = \tilde{N}(t). \quad (16)$$

Second, for the realistic situation with nonzero *b*, replication rates of B cells are changed by mutations, therefore every $m_i t$ in $(m_i t)^j$ is replaced by

$$m_i t \frac{\int_0^t m_i \exp[bih(t-t')]dt'}{\int_0^t m_i dt'} = \frac{m_i}{bih}[\exp(biht)-1], \quad (17)$$

where *t'* labels the moment that the mutation occurs. Therefore

$$N(t) = \tilde{N}(t)\sum_j P'_j(t) = \tilde{N}(t)\exp(-m_i t)\sum_{j=0}^{\infty} \frac{1}{j!}[\frac{m_i}{bih}(e^{biht}-1)]^j = \tilde{N}(t)\exp[\frac{m_i}{bih}(e^{biht}-1) - m_i t]$$
(18)

This matches Eq. (15) and more generally Eq. (14). From Eq. (18), the subpopulation of B-cells which undergo *j* mutations is

$$N_j(t) = \tilde{N}(t)P'_j(t) = \tilde{N}(t)\exp(-m_i t)\frac{1}{j!}[\frac{m_i}{bih}(e^{biht}-1)]^j, \quad (19)$$

with affinity *jih* stronger than the subpopulation without mutations. If the affinity of the initial population are all the same $X_{in}$, then $N_j(t)$ is the subpopulation with affinity $X_{in} + jih$. For *i>0*, the peak of subpopulation, i.e., largest subpopulation, is at

$$j_{peak} = \frac{m}{bih}(e^{biht}-1) \xrightarrow{t\to\infty} \frac{m}{bih}e^{biht}, \quad (20)$$

i.e., the peak moves exponentially fast in the long time limit.

From Eq. (19) a subpopulation with *j* mutations grows for two reasons, (a) fed from subpopulations with *j-1* mutations and (b) self-replication. At short enough time, *bht<<1*, the subpopulation grows $dN_j(t)/dt \approx N_{j-1}(t)m_i$ mainly for the former reason; and for large enough time, *bht>>1*, the subpopulation growth $dN_j(t)/dt \approx N_j(t)jbh$ is mainly contributed by the latter reason. Therefore, the artifact of self-replication is insignificant when time step is much shorter than the characteristic time duration *1/bh*.

**Numerical calculation for finite population.** An artifact in the above derivation is that it allows arbitrarily small $N(X,t)$, and a small subpopulation $N(X,t)<<1$ within a bin of strong affinity *X* can self-replicate rapidly. However, the B cell numbers in GCs are non-negative integers, therefore the expected number of B cells within a bin with $N(X,t)<<1$ should be zero, and cannot become the seed of a rapid growth. So the above derivations actually describe the population dynamics in the limit of infinitely large population size. It does not take into account the fact[30] that the B cell population size in a GC is no more than $10^4$.

To correct this artifact and calculate the B cell population dynamics numerically for various finite initial population sizes including the discreteness effect[7,29], we do not allow small subpopulation $N(X,t)<1$ in an affinity bin to self-replicate in our numerical calculation. Instead, it only represents an accumulative probability for the subpopulation in the bin to emerge. Our calculation is done using discrete time steps. From Eq. (15), the subpopulation which go through $\{j_i\}$ mutations $\Delta X = -ih$ between time t and $t + \Delta t$ is

$$N_{j_1 j_2 \cdots j_i \cdots}(X, t+\Delta t) = N'(X - \sum_i j_i ih, t) e^{B(X)t} e^{-m\Delta t} \prod_i \frac{[\frac{m_i}{bih}(e^{bih\Delta t} - 1)]^{j_i}}{j_i!}, \quad (21)$$

where B(X) is given by Eq. (1), and $N'(X,t)$ is the population distribution excluding the bins with less than one B cells. The population distribution after a time step is

$$N(X, t+\Delta t) = N(X,t) - N'(X,t) + \sum_{j_1=0,1,\cdots} \sum_{j_2=0,1,\cdots} \cdots \sum_{j_i=0,1,\cdots} \cdots N_{j_1 j_2 \cdots j_i \cdots}(X, t+\Delta t) \quad (22)$$

The time step $\Delta t$ is set to $0.01/bh$, much smaller the characteristic time duration $1/bh$, such that the growth of subpopulations with relatively strong affinities is dominated by mutation influx rather than self replication. We include multiple mutations in one time step. This calculation is rapid to perform even for large population sizes, and allows us to explore the parameter space.

## Results and Discussion

We calculate the total population size of B cells $N(t)$ for various initial population sizes numerically (dotted lines in Fig. 2). The total number of B cells first decreases because the initial affinity is weaker than the neutral affinity $K_a^*$. The average affinity is improved continuously (see Fig. S2 in Supporting Information) rather than abruptly, in agreement with experiments[27]. Once the average affinity reaches $K_a^*$ the population begins to increase. The picture of decrease and increase of the B cell population was observed experimentally[31] and theoretically[7], although the experimental data on GC temporal development[31] is too limited to verify the model. The lowest total population corresponds to the neutral affinity $K_a^*$; and we call it the population bottleneck, because it is the most challenging moment for the population to survive. The analytical result Eq. (14) (solid line) can describe the population size at the decreasing stage. A smaller initial population leads to a slower growth after the bottleneck (yellow). For a small enough initial population (red), the B cell population is extinct when approaching the bottleneck, and cannot recover thereafter. Therefore, the initial population size should be large enough to ensure some B cells can survive through the population bottleneck. Similarly, for a given initial population size, a weaker initial binding leads to a deeper bottleneck, and takes a longer time to recover. If the initial binding is too weak, the bottleneck will be too deep, and the population will go extinct.

The initial population is set to $10^6$ B cells in realistic calculations to include the existence of hundreds of GCs in a spleen[32] and the peak number of thousands of B cells per GC [33]. Different GCs in a spleen might not be perfectly synchronized[34]. If the B cell production rate in a spleen is limited by supply of resources, we conjecture the time that GCs start mutation and selection might vary between day 3 and day 8 or later, so that the population peaks of GCs is smeared. This agrees with the observation [35] that the total population at any moment does not exceed $2.5 \times 10^5$ B cells. The population bottleneck within a GC might also be smeared by the continued immigration of B cells from nonfollicular sites[19,36], making the bottleneck less pronounced or harder to observe directly. If all the B cells in a GC die away, the antigens are not exhausted, and it is proposed that more B cells immigrate to the GC[34], probably from GCs which have passed the bottleneck and have many B cells, although the current experiments cannot determine whether there is migration between GCs[34]. The calculations below are first performed with $10^6$ initial B cells, which is valid in the limit of fast migration between GCs in a spleen. In this case, the few GCs which by chance pass the bottleneck earlier than others may make significant contributions to the affinity improvement of the whole spleen. Then we study the case in the opposite limit, i.e. the affinity improvement of a typical GC with 3000 initial B cells, assuming no B cell migration between GCs. By comparing these two cases we will quantify the contribution of migration to affinity improvement.

We explore the parameter space to look for the optimum design of GCs that maximizes the affinity improvement (Fig. 3) in the limit of fast B cell migration between GCs ($10^6$ initial B cells). AM terminates probably due to exhaustion of available antigens[37,38] or emigration of B cells[37,39]; and the termination could be described to occur after a certain time scale, or when the B cell population size is big enough-- probably comparable to its original size,. So we assume (in Fig. 3a) termination of AM in 14 days or when the B cell population size recovers the initial value after going through the bottleneck. The affinity improvement is indicated by the total affinity A, i.e., the sum of Ka for all B cells, such that the situation with very few cells at day 14 should not be regarded as efficient improvement. The improvement of affinity in Fig. 3a is calculated for various initial binding levels and mutation rates. The highest affinity improvement is 450-fold. This optimal improvement occurs at mutation rate $m_{total} \approx 2.8/day$ (corresponding to the observed fraction of 50% daughter cells mutated), for the binding level between germline (initial) antibodies and antigens $X_{in} = X*+1 kcal/mol$ or $K_a(0)/K_a* = \exp(-1kcal/mol/kT) = 0.18$, and selection strength $b = 0.7/day(kcal/mol)$ (see Fig. S3 and S4 in Supporting Information). This result agrees with several experimental observations. First, the affinity improvement agrees with the observation of ~100-fold[11,12,13,14] improvement. Second, the theoretically optimal value to provide maximal affinity improvement agrees with the observed *in vivo* somatic hypermutation rate[9]. Third, the improvement of affinity corresponds to ln(450)kT=3.6kcal/mol of free energy improvement. Combining with the typical affinity improvement $Y \approx 0.4$kcal/mol of an affinity improving mutation, we can estimate that a final B cell contains 3.6/0.4=9 mutations in their V regions of Ig genes, in agreement with the observed ~9 mutations per Ig gene[9,40,41,42,43]. From the definition $b = 1/(kT\tau)$, the optimal selection strength b=0.7/day/(kcal/mol) corresponds to an optimal time of a recycling round $\tau = 1/(kTb) \approx 2.4$ day, compatible with the earlier model[7].

Fig. 3b helps to reveal the design principles of GCs, where the affinity improvement is calculated when the B cell population size recovers the initial value after going through the bottleneck, no matter how long it takes. In general, the affinity improvement is not effective for too strong initial bindings, which results in shallow or no bottlenecks and rapid population recoveries, hastily terminating the AM before accumulating adequate improvements. As the initial binding becomes weaker, the population bottleneck is deeper, and the affinity improvement is more effective, but takes a longer time. The improvement shown in Fig. 3b could even exceed 1000-fold for a weak initial binding $X_{in} = X*+1.5 kcal/mol$, although such improvements take much longer than 14 days and would be interrupted in the calculation with fixed AM time as shown in Fig. 3a. If the initial binding becomes even weaker, the population bottleneck is so deep that the whole population goes extinct and no longer recovers (grey scales in Fig. 3b obtained numerically). Therefore, for a given mutation rate, the most effective affinity improvement occurs when initial binding level is near the critical value (red in Fig. 3b), where the population can barely survive through the population bottleneck, i.e. only a few GCs can survive. A somewhat stronger initial binding improves affinities less effectively, but it takes a shorter time for the AM to finish, and faster AM is advantageous. Taking all these into consideration, in the optimal design of GCs, the most commonly appeared initial bindings should be adjusted to be somewhat stronger than the critical value, so that affinities are improved effectively, timely, and safely. This picture agrees with the observed dependence of B cell fate on initial antibody-antigen binding level or antigen density[44] [45], where too strong initial bindings (beyond Ka* in our model) do not lead to GC formation, moderate initial bindings (e.g. $X_{in} = X*+0.5 kcal/mol$ in our model) result in GC response which finishes quickly, while weaker initial affinities (e.g. $X_{in} = X*+1 kcal/mol$ in our model) result in tempered GCs. To achieve the optimal design, the $Ka^*$ values might have been adjusted in evolution by tuning the antigen density or modulating the diversity of the germline pool achieved through somatic recombination. When mutation rate is so high that 80% or more daughter cells are mutated, the lethal mutations preclude sustainable replications of B cells, i.e., the population growth rate in Eq. (1) becomes negative for any affinity. This lethal mutagenesis region is shown in black in Fig. 3b.

Exploration of the parameter space helps us sketch some general design principles of the GCs. First, for the affinity to be improved most effectively, the B cell number should first decrease to reach a population bottleneck and then increase. A weaker initial binding results in a higher affinity improvement, unless the population recovery after the bottleneck becomes too slow, or even takes forever in the extinct case. Second, the seemingly high mutation rate is actually set to optimize the success rate of AM. On one hand, the optimal mutation rate, in agreement with somatic hypermutation *in vivo*, is quite high because the improvement of affinity comes only from mutations. On the other hand, if the mutation rate gets higher, the large number of lethal mutations will spoil the cell replications. Third, we expect that the selection strength b is the optimal value b=0.7/day/(kcal/mol), which sets an important guide to future simulations of GCs. Indeed, a large enough b ensures adequate preference for stronger affinity B cells in selection; but if b is too large, the initial weak binding B cells vanish quickly before enough beneficial mutations can accumulate.

B cell migration between GCs could be beneficial to the AM, as indicated by the calculated affinity improvement of a typical GC with 3000 initial B cells in the limit of no migration (Fig. 4). A stronger initial binding $X_{in} = X*+0.5 kcal/mol$ is needed to ensure a typical GC to survive through the bottleneck; and the optimal affinity improvement becomes 70-fold for the GC, still consistent with the experimental value (of the order ~100-fold[11,12,13,14]). This is achieved in 16 days, only slightly longer than two weeks, therefore separate diagrams like in Fig. 3a and b is not necessary. The maximal improvement in this case is ~6-7 times lower than the result in the limit of fast migration, where a factor of ~3 comes from the change in initial binding level, and a factor of ~2 comes from the difference between the average of all GCs and the "typical" or median individual GC. In other words, fast B cell migration between GCs could enhance the affinity improvement by a factor of about 6-7 in our model.

Our model is consistent with the ''all or none[8]'' phenomenon observed in experiments[19,46,47], i.e., the fraction of strong affinity cells in a GC is most likely either very high or very low, where the strong affinity B cells are characterized by a certain key mutation[46,47] or a unique piece of Ig gene sequence[19]. Presumably the GCs dominated by strong affinity mutants have gone through the population bottleneck, and this phenomenon is the most pronounced in the case without B cell migration between GCs. Indeed, it is possible for different individual GCs to fall into two categories according to whether they have passed the population bottleneck, and the choice of category for a GC depends on both the initial binding level and stochastic effect. Moreover, if many GCs in a spleen start from the same initial condition but each has a random starting time, then the fractions of strong affinity B cells in the GCs at a given moment is expected to be distributed as in Fig. 5, where affinities beyond Ka* are defined as strong. Fig. 5 is obtained as follows. During the development of a typical GC (Fig. 2 and S2), we can track the fraction $F(t)$ of strong affinity B cells in the GC all the time. Observing the whole ensemble of GCs at a given moment is equivalent to observing a single GC at many arbitrary moments. Therefore, we transform $F(t)$ into $t(F)$, and the distribution is $P(F) \sim dt(F)/dF$ up to a normalization factor. As we see in Fig. 5, the probabilities to observe high or low fraction values are significantly larger than that of intermediate fraction values. Therefore, our model is consistent with the ''all or none'' phenomenon. If B cell diffusions between GCs exist, this phenomenon is somewhat smeared.

We obtain mutation effects $W(\Delta X)$ from the PINT database which presents the data on interactions between all kinds of proteins[26] because we do not have adequate real data of affinity change upon mutations of Ig. The existing analysis of Ig sequence data in the immune response of PhOx and NP[8] is consistent with the distribution we obtain from PINT database. Indeed, 3.2% (for PhOx) or 1.0% (for NP) of the affinity affecting mutations could improve affinity strongly (10-fold), while in our model 4.9% improves affinity and 1.4% improves affinity strongly (5-fold). Our assumption that the distribution of affinity change $W(\Delta X)$ is independent on the affinity might not hold for all affinities, although our model only requires this assumption to be valid just for a range 2-4 kcal/mol of affinities.

What are the major differences between our model and the earlier theoretical models? First, the affinity improvement we achieve is higher than the earlier theoretical models, which resolves the discrepancy between theory and experiment. If we examine Oprea-Perelson model[7] in our framework, the reason that the earlier model did not achieve higher improvements might be that it used a smaller selection strength b than the optimal value. Indeed, the choice to enter a recycle in the Oprea-Perelson model is stochastic, so the chance for weak affinity B cells to survive is enhanced by choosing not to enter a cycle and therefore avoiding selection. A larger selection strength b brings a high risk of AM failure, but provides a higher affinity improvement, and this can be achieved if all B cells experience similar number of recycles rounds. The low optimal value of mutation rate in the earlier model might result from the small selection strength. Second, the major simplification we make is to use Eq. (1) to replace the binding kinetics between Ig and antigens, salvation, recycling, the distinction between different B cell phenotypes such as centroblasts and centrocytes, and the separation of dark and light zones, which were considered in details in earlier theoretical studies[7,8,37,38,39,48,49]. We assume that competition for antigen and T-cell help effectively results in the dependence of death rate of B cells on their affinity to an antigen, and we provided an argument that such dependence should be linear in binding free energy of Ig-antigen interaction. We expect that this simplification captures the key factor determining the AM. Such a coarse-grained description reduces the parameter space greatly, making it possible to search the whole parameter space for optimal design of GC. With this basic picture, we can build more detailed simulations to reproduce the complete procedures of AM.

## Acknowledgements

We are grateful to S. Wylie for helpful discussions, and A. Perelson for drawing our attention to the improvement of affinity. This work is supported by NIH.

# References

1. MacLennan IC (1994) Germinal centers. Annu Rev Immunol 12: 117-139.
2. Rajewsky K (1996) Clonal selection and learning in the antibody system. Nature 381: 751-758.
3. Allen CD, Okada T, Cyster JG (2007) Germinal-center organization and cellular dynamics. Immunity 27: 190-202.
4. Schwickert TA, Lindquist RL, Shakhar G, Livshits G, Skokos D, et al. (2007) In vivo imaging of germinal centres reveals a dynamic open structure. Nature 446: 83-87.
5. Neuberger MS (2002) Novartis Medal Lecture. Antibodies: a paradigm for the evolution of molecular recognition. Biochem Soc Trans 30: 341-350.
6. Kepler TB, Perelson AS (1993) Somatic hypermutation in B cells: an optimal control treatment. J Theor Biol 164: 37-64.
7. Oprea M, Perelson AS (1997) Somatic mutation leads to efficient affinity maturation when centrocytes recycle back to centroblasts. J Immunol 158: 5155-5162.
8. Kleinstein SH, Singh JP (2001) Toward quantitative simulation of germinal center dynamics: biological and modeling insights from experimental validation. J Theor Biol 211: 253-275.
9. Berek C, Milstein C (1987) Mutation drift and repertoire shift in the maturation of the immune response. Immunol Rev 96: 23-41.
10. Kepler TB, Perelson AS (1995) Modeling and optimization of populations subject to time-dependent mutation. Proc Natl Acad Sci U S A 92: 8219-8223.
11. Torigoe H, Nakayama T, Imazato M, Shimada I, Arata Y, et al. (1995) The affinity maturation of anti-4-hydroxy-3-nitrophenylacetyl mouse monoclonal antibody. A calorimetric study of the antigen-antibody interaction. J Biol Chem 270: 22218-22222.
12. Sharon J (1990) Structural correlates of high antibody affinity: three engineered amino acid substitutions can increase the affinity of an anti-p-azophenylarsonate antibody 200-fold. Proc Natl Acad Sci U S A 87: 4814-4817.
13. Ulrich HD, Mundorff E, Santarsiero BD, Driggers EM, Stevens RC, et al. (1997) The interplay between binding energy and catalysis in the evolution of a catalytic antibody. Nature 389: 271-275.
14. Yang PL, Schultz PG (1999) Mutational analysis of the affinity maturation of antibody 48G7. J Mol Biol 294: 1191-1201.
15. Kroese FG, Wubbena AS, Seijen HG, Nieuwenhuis P (1987) Germinal centers develop oligoclonally. Eur J Immunol 17: 1069-1072.
16. Coffey F, Alabyev B, Manser T (2009) Initial clonal expansion of germinal center B cells takes place at the perimeter of follicles. Immunity 30: 599-609.
17. McHeyzer-Williams MG, McLean MJ, Lalor PA, Nossal GJ (1993) Antigen-driven B cell differentiation in vivo. J Exp Med 178: 295-307.
18. Han S, Zheng B, Dal Porto J, Kelsoe G (1995) In situ studies of the primary immune response to (4-hydroxy-3-nitrophenyl)acetyl. IV. Affinity-dependent, antigen-driven B cell apoptosis in germinal centers as a mechanism for maintaining self-tolerance. J Exp Med 182: 1635-1644.

19. Jacob J, Przylepa J, Miller C, Kelsoe G (1993) In situ studies of the primary immune response to (4-hydroxy-3-nitrophenyl)acetyl. III. The kinetics of V region mutation and selection in germinal center B cells. J Exp Med 178: 1293-1307.
20. Pascual V, Liu YJ, Magalski A, de Bouteiller O, Banchereau J, et al. (1994) Analysis of somatic mutation in five B cell subsets of human tonsil. J Exp Med 180: 329-339.
21. Allen CD, Okada T, Tang HL, Cyster JG (2007) Imaging of germinal center selection events during affinity maturation. Science 315: 528-531.
22. Hanna MG, Jr. (1964) An Autoradiographic Study of the Germinal Center in Spleen White Pulp During Early Intervals of the Immune Response. Lab Invest 13: 95-104.
23. Zhang J, MacLennan IC, Liu YJ, Lane PJ (1988) Is rapid proliferation in B centroblasts linked to somatic mutation in memory B cell clones? Immunol Lett 18: 297-299.
24. Shannon M, Mehr R (1999) Reconciling repertoire shift with affinity maturation: the role of deleterious mutations. J Immunol 162: 3950-3956.
25. Shlomchik MJ, Watts P, Weigert MG, Litwin S (1998) Clone: a Monte-Carlo computer simulation of B cell clonal expansion, somatic mutation, and antigen-driven selection. Curr Top Microbiol Immunol 229: 173-197.
26. Kumar MD, Gromiha MM (2006) PINT: Protein-protein Interactions Thermodynamic Database. Nucleic Acids Res 34: D195-198.
27. Kocks C, Rajewsky K (1988) Stepwise intraclonal maturation of antibody affinity through somatic hypermutation. Proc Natl Acad Sci U S A 85: 8206-8210.
28. Batista FD, Neuberger MS (1998) Affinity dependence of the B cell response to antigen: a threshold, a ceiling, and the importance of off-rate. Immunity 8: 751-759.
29. Tsimring LS, Levine H, Kessler DA (1996) RNA virus evolution via a fitness-space model. Phys Rev Lett 76: 4440-4443.
30. Kuppers R, Zhao M, Hansmann ML, Rajewsky K (1993) Tracing B cell development in human germinal centres by molecular analysis of single cells picked from histological sections. Embo J 12: 4955-4967.
31. Liu YJ, Zhang J, Lane PJ, Chan EY, MacLennan IC (1991) Sites of specific B cell activation in primary and secondary responses to T cell-dependent and T cell-independent antigens. Eur J Immunol 21: 2951-2962.
32. Jacob J, Kassir R, Kelsoe G (1991) In situ studies of the primary immune response to (4-hydroxy-3-nitrophenyl)acetyl. I. The architecture and dynamics of responding cell populations. J Exp Med 173: 1165-1175.
33. Smith KG, Light A, Nossal GJ, Tarlinton DM (1997) The extent of affinity maturation differs between the memory and antibody-forming cell compartments in the primary immune response. Embo J 16: 2996-3006.
34. Or-Guil M, Wittenbrink N, Weiser AA, Schuchhardt J (2007) Recirculation of germinal center B cells: a multilevel selection strategy for antibody maturation. Immunol Rev 216: 130-141.
35. Shahaf G, Barak M, Zuckerman NS, Swerdlin N, Gorfine M, et al. (2008) Antigen-driven selection in germinal centers as reflected by the shape characteristics of immunoglobulin gene lineage trees: a large-scale simulation study. J Theor Biol 255: 210-222.
36. Jacob J, Kelsoe G (1992) In situ studies of the primary immune response to (4-hydroxy-3-nitrophenyl)acetyl. II. A common clonal origin for periarteriolar lymphoid sheath-associated foci and germinal centers. J Exp Med 176: 679-687.
37. Iber D, Maini PK (2002) A mathematical model for germinal centre kinetics and affinity maturation. J Theor Biol 219: 153-175.
38. Kesmir C, De Boer RJ (1999) A mathematical model on germinal center kinetics and termination. J Immunol 163: 2463-2469.
39. Meyer-Hermann M, Deutsch A, Or-Guil M (2001) Recycling probability and dynamical properties of germinal center reactions. J Theor Biol 210: 265-285.
40. Wedemayer GJ, Patten PA, Wang LH, Schultz PG, Stevens RC (1997) Structural insights into the evolution of an antibody combining site. Science 276: 1665-1669.
41. Berek C, Berger A, Apel M (1991) Maturation of the immune response in germinal centers. Cell 67: 1121-1129.
42. Sharon J, Gefter ML, Wysocki LJ, Margolies MN (1989) Recurrent somatic mutations in mouse antibodies to p-azophenylarsonate increase affinity for hapten. J Immunol 142: 596-601.
43. Siekevitz M, Kocks C, Rajewsky K, Dildrop R (1987) Analysis of somatic mutation and class switching in naive and memory B cells generating adoptive primary and secondary responses. Cell 48: 757-770.
44. O'Connor BP, Vogel LA, Zhang W, Loo W, Shnider D, et al. (2006) Imprinting the fate of antigen-reactive B cells through the affinity of the B cell receptor. J Immunol 177: 7723-7732.
45. Paus D, Phan TG, Chan TD, Gardam S, Basten A, et al. (2006) Antigen recognition strength regulates the choice between extrafollicular plasma cell and germinal center B cell differentiation. J Exp Med 203: 1081-1091.
46. Ziegner M, Steinhauser G, Berek C (1994) Development of antibody diversity in single germinal centers: selective expansion of high-affinity variants. Eur J Immunol 24: 2393-2400.
47. Radmacher MD, Kelsoe G, Kepler TB (1998) Predicted and inferred waiting times for key mutations in the germinal centre reaction: evidence for stochasticity in selection. Immunol Cell Biol 76: 373-381.
48. Meyer-Hermann ME, Maini PK, Iber D (2006) An analysis of B cell selection mechanisms in germinal centers. Math Med Biol 23: 255-277.
49. Celada F, Seiden PE (1996) Affinity maturation and hypermutation in a simulation of the humoral immune response. Eur J Immunol 26: 1350-1358.

# Figure Captions

**Figure 1:** Histogram of affinity improvement upon single mutations from the PINT database. The silent or lethal mutation is not included in the figure. Here the bin size is $h=0.5$ kcal/mol (equivalent to 2.3-fold in Ka), and the unit of $W$ is $(kcal/mol)^{-1}$. Only 4.9% of the affinity affecting mutations could improve affinity.

**Figure 2**: Total population as a function of time for various initial population sizes, starting from germline (initial) Ig-antigen binding level $X_{in} = X^* + 1 kcal/mol$ or $Ka_{in}/Ka^* = 0.18$, with $b = 0.7/day(kcal/mol)$ and $m_{total} = 2.8/day$. Solid line: exact analytical result for infinite population size, Eq. (14). Dotted lines: numerical results for initial population sizes $N_0 = 10^7$ (green), $10^5$ (yellow), and $10^2$ (red) respectively. The population in red goes extinct at the bottleneck.

**Figure 3**: The improvement of affinity for the whole spleen including many GCs in the limit of rapid B cell migration between GCs. The improvement of affinity (in Ka) is shown in color code, as a function of mutation rate and initial binding level, assuming AM is terminated (a) either after 14 days or when the population recovers the initial size ($10^6$ B cells) after going through the bottleneck, or (b) when the population recovers the initial size after going through the bottleneck, no matter how long it takes. (a): The optimal improvement of A (sum of Ka over all B cells) occurs when about 50% daughter cells are mutated at divisions. (b): The grey scale shows the probability for the whole population to survive through the bottleneck. The black region at high mutation rate indicates lethal mutagenesis where there are too many lethal mutations for B cell population to increase.

**Figure 4**: The improvement of affinity for an isolated GC, i.e., in the limit of no B cell migration between GCs. The improvement of affinity (in Ka) is shown in the same color code as in Fig. 3a, assuming AM is terminated when the population recovers the initial size (3000 B cells). The optimal improvement of affinity occurs when about 60% daughter cells are mutated at divisions, and takes 16 days. The grey scale shows the probability for a GC to survive through the bottleneck.

**Figure 5**: The distribution of F, fraction of strong affinity B cells, for many GCs which follow similar development patterns but each starts from a random time, is consistent with the "all or none" phenomenon. Every GC has initial binding level $X_{in} = X^* + 0.5 kcal/mol$, 50% mutated daughter cells, and selection strength $b = 0.7/day(kcal/mol)$. The calculation is terminated as F reaches 85%, when the number of B cells in the GC recovers the initial value (3000).

Figure 1

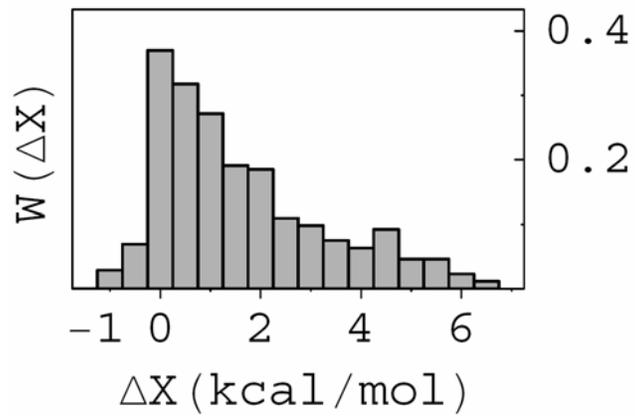

Figure 2

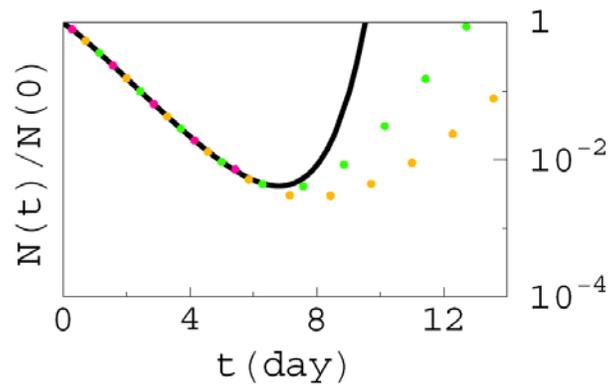

Figure 3

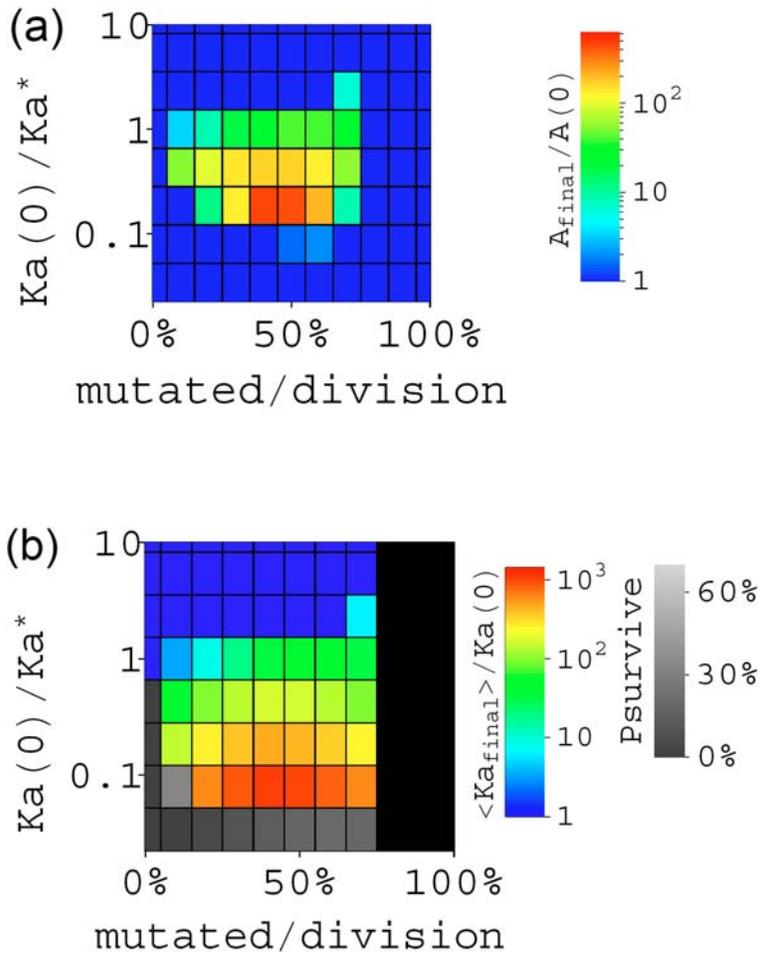

Figure 4

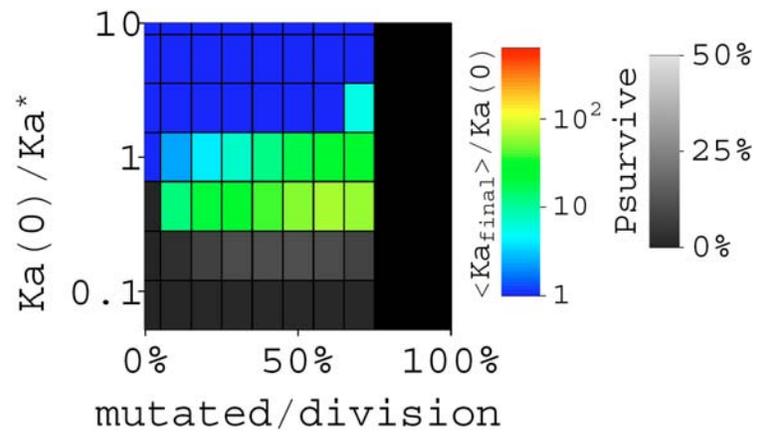

Figure 5

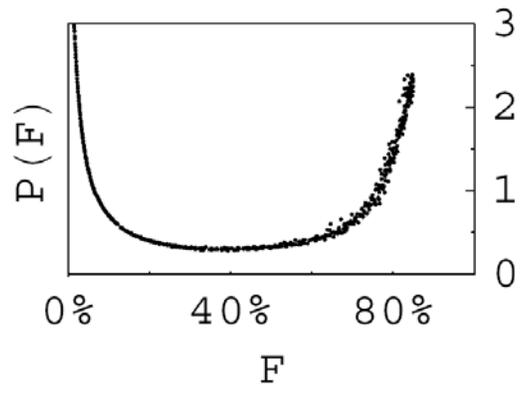

## Supplementary Information

**Affinity change upon mutations:** From the PINT (protein interaction) database[1], the change of affinity upon mutations does not have obvious correlation with the affinity before mutations (Fig. S1). The fraction of beneficial mutations has no significant dependence on the affinity, especially for the typical range of affinity maturation $10^6 M^{-1} < Ka < 10^9 M^{-1}$ or $-12 kcal/mol < X < -8 kcal/mol$. The detrimental mutations in Fig. 1 can be fitted by $W(\Delta X) = 0.48 \exp[-\Delta X /(1.87 kcal/mol)]$, while the data of beneficial mutations can be fitted by $W(\Delta X) = 0.18 \exp[-\Delta X /(0.43 kcal/mol)]$.

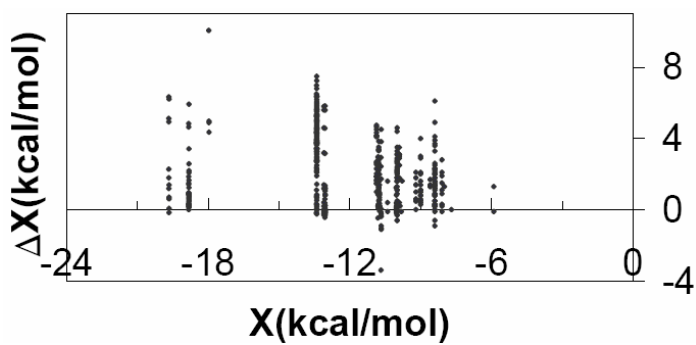

Fig. S1: The scatter plot of affinity X and the change of affinity $\Delta X$ from PINT database, which does not show significant correlation.

**Survival probability:** If the initial B cell population is too small, the population might not pass through the bottleneck, but rather becomes extinct. To calculate the survival probability of the whole population numerically, we start from an initial B cell with weak affinity $X$, corresponding to a death rate $|B(X)|$ larger than the replication rate $r - 0.3 m_{total}$. The initial cell and its descendants could either die or replication and/or mutate with some well defined probabilities. By collecting the probability distribution of each cell's fate, we numerically calculate the probability for some descendants to reach a strong affinity $X^* - 10h$. In this way we find the probability $p$ for some descendants of the initial B cell with affinity X to eventually survive, and use $p$ to find the expected probability for the system of $10^6$ initial B cells to survive rather than become extinct.

**Gradual improvement of affinity:** We can monitor the distribution of affinity at different moments (Fig. S2). The population starts from identical weak affinity B cells $X_{in} = X^* + 1 kcal/mol$ or $Ka_{in}/Ka^* = 0.18$, and has an initial population decrease; but B cells with stronger affinities gradually emerge, and become dominant due to selection, and the population size grows thereafter. The continuous improvement of affinity is supported by experiments[2].

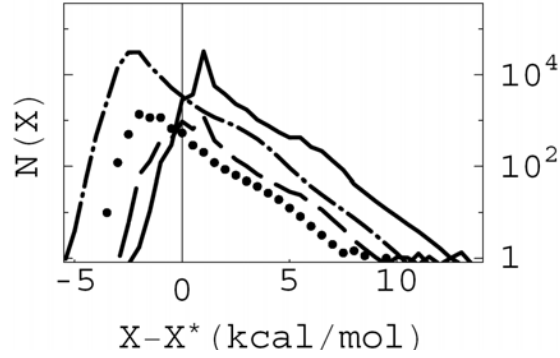

Figure S2: Affinity distribution in the population $N(X,t)$ at t=3 (solid), 6 (dashed), 9 (dotted), and 12 (dash-dotted) days, starting from germline affinity $X_{in} = X^* + 1 kcal/mol$ or $Ka_{in}/Ka^* = 0.18$, with initial population $N_0 = 10^6$, selection strength $b = 0.7/day(kcal/mol)$ and effective mutation rate $m_{total} = 2.8/day$. The average affinity improves with time; while population size shrinks then grows.

**Affinity Improvement as a function of selection strength:** We look for the optimal values of mutation rate $m$, selection strength $b$ and initial binding level $X_{in} - X^*$ that maximize affinity improvement for the case of fast B cell migration between GCs (i.e. $N_0 = 10^6$ initial B cells). The variables in Fig. 3a are $m$ and $X_{in} - X^*$, and here we plot the improvement of affinity in the plane of $b$ and $m$ (Fig. S3), and in the plane of $b$ and $X_{in} - X^*$ (Fig. S4). At each given value of $b$ and $m$ in Fig. S3, different initial affinities $X_{in} - X^* = 0, 0.5, 1, 1.5$, and $2 kcal/mol$ are calculated, and the initial affinity resulting in the strongest affinity improvement is chosen. To maximize the affinity improvement, the optimal mutation rate is at 50% daughter cells mutated at each replication, and the optimal selection strength is b=0.7/day/(kcal/mol).

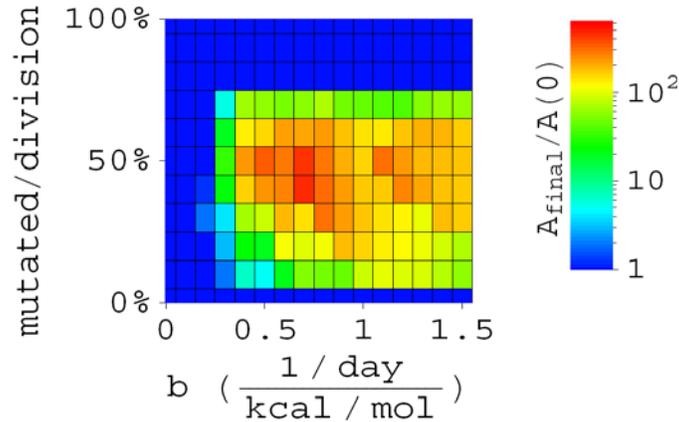

Fig. S3: Optimization of $b$ and $m$. The color indicates the improvement of total affinity. Here different initial affinities are tried for each mutation rate and b, and the one which gives largest affinity improvement is chosen. b=1.2/day/(kcal/mol) is the global optimal selection strength. A minor local peak at b=1.2/day/(kcal/mol) might be an artifact due to discrete (rather than continuous) choices of initial affinity values.

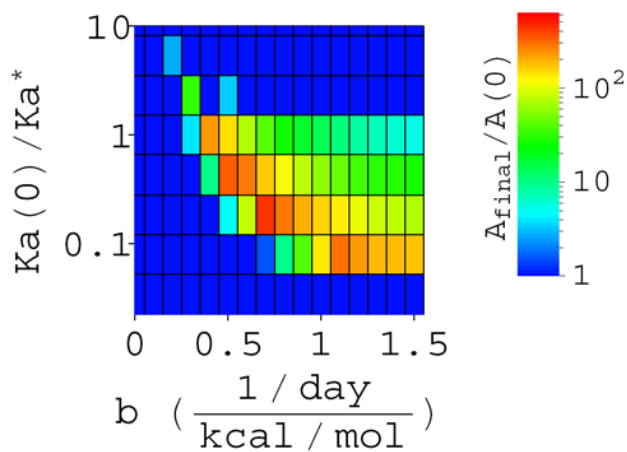

Fig. S4: The improvement of total affinity as a function of selection strength and initial affinity. Here mutation rate is chosen as the optimum value, i.e. m=0.55/day/gene or 50% mutated daughter cells.

**References**

1. Kumar MD, Gromiha MM (2006) PINT: Protein-protein Interactions Thermodynamic Database. Nucleic Acids Res 34: D195-198.
2. Kocks C, Rajewsky K (1988) Stepwise intraclonal maturation of antibody affinity through somatic hypermutation. Proc Natl Acad Sci U S A 85: 8206-8210.